# Interplay of structure, magnetism, and magneto-thermal effects in Gadolinium-based intermetallic compound

Ravinder Kumar[1,†], Arrab Ali Maz[1,†], Satyendra Kumar Mishra[3] and Sachin Gupta[1,2,*]

[1]*Department of Physics, Bennett University, Greater Noida 201310, India*

[2]*Centre of Excellence in Nanosensors and Nanomedicine, Bennett University, Greater Noida 201310, India*

[3]*Space and Resilient Communications and Systems (SRCOM), Center Technologic de Telecomunicacions de Catalunya (CTTC), Avinguda Carl Friedrich Gauss, 11, 08860 Castelldefels, Spain*

*Correspondence: sachin.gupta@bennett.edu.in

† both authors contributed equally.

**Abstract:**

We study structural, magnetic and magneto-thermal properties of GdRhIn compound. The room temperature X-ray diffraction measurements show hexagonal crystal structure. Temperature and field dependence of magnetization suggest two magnetic transitions – antiferromagnetic to ferromagnetic at 16 K and ferromagnetic to paramagnetic at 34. The heat capacity measurements confirm both the magnetic transitions in GdRhIn. The magnetization data was used to calculate isothermal magnetic entropy change and refrigerant capacity in GdRhIn, which was found to be 10.3 J/Kg-K for the field change of 70 kOe and 282 J/Kg for the field change of 50 kOe, respectively. The large magnetocaloric effect in GdRhIn suggests that the material could be used for magnetic refrigeration at low temperatures.

# 1. Introduction

Magnetism originated from unpaired 4*f* shell electrons in rare earths make them interesting for fundamental physics as well as for application purposes. When these materials are combined with the transition metals, they show outstanding properties emerging from rare earth and transition metal sublattices. Rare earth intermetallics are being investigated for various applications such as permanent magnets, magnetic sensors, magnetic refrigeration, bio-medical and spintronics [1–8]. Among rare earth intermetallics, *RTX* (*R*= rare earth, *T*= transition metal, *X*= main group element) compounds were found to show many unusual thermal and magnetic properties, which have drawn significant attention from scientists and engineers [1,2,4]. Physical properties of various *RTX* compounds were studied in detail to corelate their structural, magnetic and magnetocaloric properties, some are referenced in Refs. [9–27]. However, there are still some materials for which their magnetic and other properties remain unexplored. Among others, *R*RhIn series is one of them. *R*RhIn have been reported to crystallize in two crystal structures - *R*RhIn with *R*=La-Nd, Sm, Gd-Tm in ZrNiAl type hexagonal and *R*=Eu, Yb, Lu in TiNiSi type orthorhombic crystal structure, respectively [20,28–31]. EuRhIn and GdRhIn in this series were found to show ferromagnetic (FM) behavior below 22 and 34 K, respectively [32,33]. Though most of members in this series lack detailed study on magnetic and related properties, in this paper, we discuss combined structural, magnetic and magneto-thermal properties of GdRhIn compound.

We synthesized GdRhIn compound using arc melt technique and studied structural, magnetic, and magneto-thermal properties. Cystal structure refinement confirms hexagonal crystal structure in GdRhIn. The magnetic measurements show double magnetic transitions at low temperatures. To confirm it, we carried out the heat capacity measurements. To know the potential of this material for magnetic refrigeration applications, we calculated magnetocaloric effect (MCE) from magnetization data observed that GdRhIn shows large MCE.

## 2. Experimental Details

The arc melting technique was used to melt constituent elements Gd, Rh and In. The atomic purity for each element was better than 99.9%. To ensure better homogeneity the formed ingot was flipped multiple times and re-melted. The arc melted sample was sealed in evacuated ($10^{-6}$ torr) quartz tube and annealed in a furnace for 7 days at 800 ˚C. The X-ray diffraction measurement of powder sample was performed at X'PERT PRO diffractometer using Cu K$\alpha$ radiation at room temperature. Magnetic measurements were performed at Vibrating Sample Magnetometer (VSM) attached to a Physical Property Measurement System (Quantum Design, PPMS-6500). PPMS was also used to measure the heat capacity of the sample using thermal relaxation method.

## 3. Results and discussion

Figure 1 shows powder XRD pattern recorded at room temperature for GdRhIn compound. The obtained XRD pattern was analyzed with Rietveld refinement using FullProf Suite software [34]. The Rietveld refinement of XRD pattern confirms that GdRhIn crystallizes in Fe$_2$P type hexagonal crystal structure with space group P-62m (Space group #189). The lattice parameters obtained for the refinement are $a = b = 7.53$ Å and $c = 3.92$ Å for GdRhIn compound

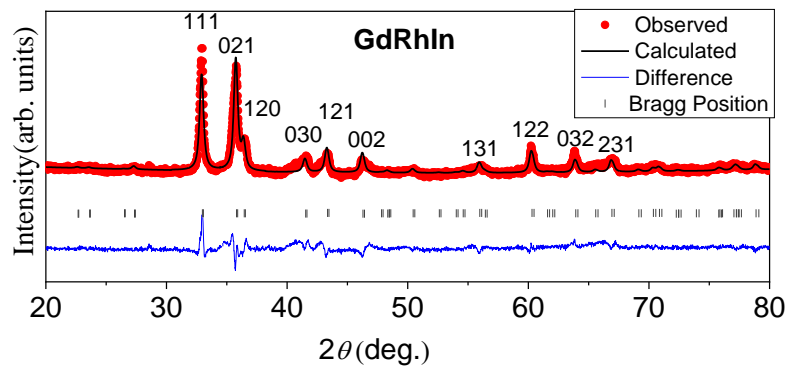

**Figure 1.** Powder X-ray diffraction pattern for GdRhIn compound. The bottom plot shows the difference between the theoretically calculated and experimentally observed data. The high intensity peaks are indexed with the Bragg notation.

The temperature, $T$ and magnetic field, $H$ dependence of magnetization, $M$ for GdRhIn is shown in Fig. 2. Fig. 2(a) shows the temperature dependence of magnetization under zero field cooled (ZFC) and field cooled (FC) configurations at applied field of 500 Oe. The difference between ZFC and FC modes is that, in the first case, the sample is cooled in the absence of any magnetic field, while in the latter case, it is cooled in the presence of a magnetic field. The magnetization in both the cases is recorded in the presence of a magnetic field as the sample is heated. It can be noted from Fig. 2(a) that on decreasing the temperature, the GdRhIn shows paramagnetic to ferromagnetic transition at 34 K. At 16 K, the magnetization shows a cusp, suggesting antiferromagnetic transition. At low temperatures, the difference in magnetization for ZFC and FC is attributed due to thermomagnetic irreversibility as seen in many *RTX* compounds [35]. The Curie-Weiss fit to susceptibility data yields the values of paramagnetic Curie temperature ($\theta_p$) and effective magnetic moment ($\mu_{eff}$) to be 24.3 K and 8 $\mu_B/Gd^{3+}$, respectively. It is worth noting that the $\theta_p$ value obtained with positive sign suggests that ferromagnetic correlations among the magnetic moments are strong. The observed value of $\mu_{eff}$ is very close to theoretical expected value of 7.94 $\mu_B/Gd^{3+}$ for Gadolinium ion. Fig. 2(b) shows the field dependence of magnetization measured at different temperatures. At the field of 90 kOe, the magnetization in GdRhIn shows clear saturation (main plot). The inset in Fig. 2(b) shows magnified view of magnetic field dependence of magnetization. It can be noted from the inset that at low magnetic fields, the magnetization increases with increasing temperature up to 16 K, indicating antiferromagnetic transition. At higher magnetic fields, the magnetization decreases with temperature, indicating ferromagnetic transition below 34 K. The saturation moment for field value of 90 kOe and at 5 K was found to be 6.9 $\mu_B$, which is very close to theoretical value of 7 $\mu_B/Gd^{3+}$.

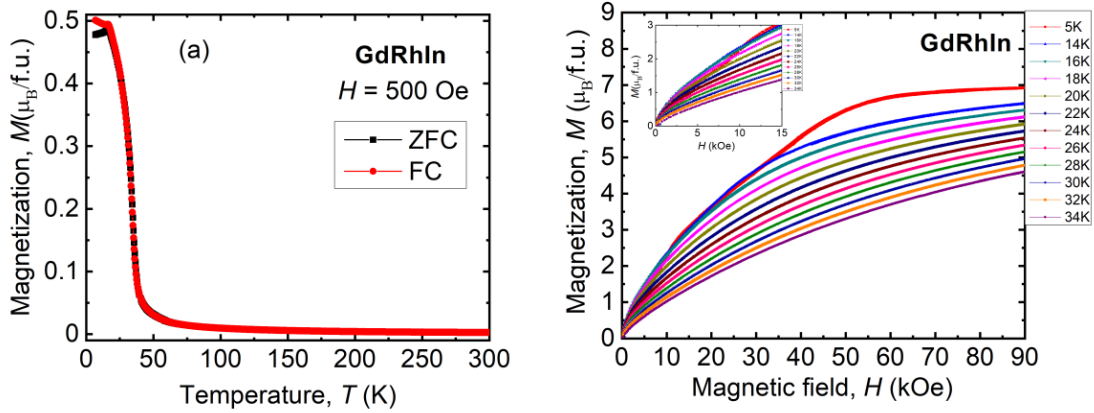

**Figure 2.** (a) The temperature, $T$ dependence of magnetization, $M$ at magnetic field of 500 Oe recorded under ZFC and FC modes. (b) The field, $H$ dependence of magnetization, recorded at different temperatures for GdRhIn compound. The inset shows magnified view of M vs. H plot at various temperatures.

In order to know more about thermal and magnetic behavior of GdRhIn, the heat capacity, $C$ measurement was performed at zero magnetic field as shown in Fig. 3. The temperature dependence of the heat capacity shows two anomalies corresponding to the two magnetic transitions as observed in magnetization data. The $\lambda$- type peak is observed around 16 K due to antiferromagnetic transition, while slight change in the slope can be seen around 34 K which corresponds to ferromagnetic transition in GdRhIn.

The temperature dependence of the heat capacity (lattice and electronic) in low temperature regime can be expresses by following equation:

$$C = \gamma T + \beta T^3 \qquad (1)$$

Where $\gamma$ and $\beta$ are the coefficient of electronic heat capacity and thermal expansion, respectively. The Debye temperature, $\theta_D$ can be deduced from the following relation [36]

$$\theta_D = \sqrt[3]{12\pi^4 R/5\beta} \cong \sqrt[3]{1944/\beta} \qquad (2)$$

The inset in Fig. 3 shows a fit of equation (1) to experimental heat capacity data at low temperatures. The $\gamma$ and $\beta$ values estimated from the fit are found to be 0.79 and 0.0006, respectively. Due to the low ordering temperature in GdRhIn, the value of $\gamma$ could be deviated from a reasonable value. The Debye temperature, $\theta_D$ estimated using equation (2) is found to be 148 K, which is close to values observed in other similar *RTX* compounds [35,37,38].

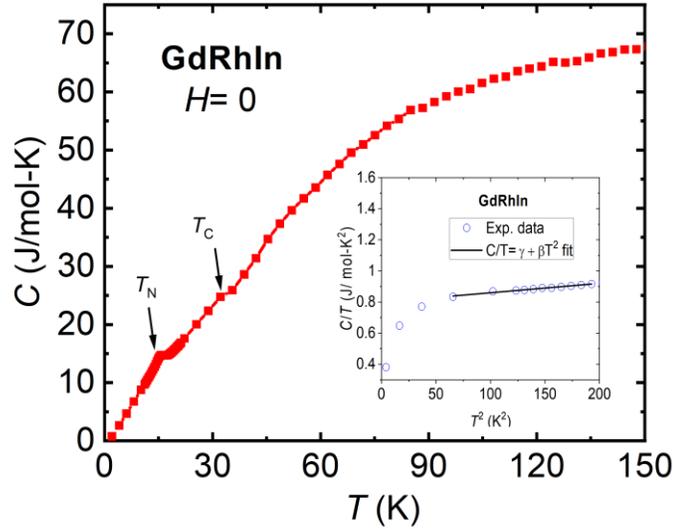

**Figure 3**. The temperature dependence of the heat capacity, *C* measured at zero magnetic field. The arrows are a guide to magnetic transition temperatures. The inset shows fit to heat capacity data in low temperature regime.

Magnetocaloric effect (MCE) is an intrinsic property of magnetic material, which makes material heat up or cool down on the application or removal of the magnetic field. MCE can be measured in terms of change in isothermal magnetic entropy, $\Delta S_M$ and/or adiabatic temperature, $\Delta T_{ad}$. To know the potential of GdRhIn for magnetic refrigeration applications, the change in the isothermal magnetic entropy, $\Delta S_M$ was calculated from the magnetization, *M* (*H*, *T*) data using the following Maxwell's relation

$$\Delta S_M = \int_0^H \left[\frac{\partial M}{\partial T}\right]_H dH \qquad (3)$$

The above relation was simplified in following equation, which can be used for calculating isothermal magnetic entropy change [39]

$$\Delta S_M = \sum_i \frac{M_{i+1} - M_i}{T_{i+1} - T_i} \Delta H_i \qquad (4)$$

Where, $M_{i+1}$ and $M_i$ represent the magnetization at temperatures $T_{i+1}$ and $T_i$, respectively in the field $H_i$.

Fig. 4(a) shows the temperature dependence of $\Delta S_M$ as a function of magnetic field for GdRhIn compound. It can be observed from the plot that the value of $\Delta S_M$ increases with field and shows a value of 10.3 J/Kg-K for the field change of 70 kOe. The obtained value of $\Delta S_M$ in GdRhIn is higher than many other Gd-based materials [40–43]. The temperature dependence of $\Delta S_M$ shows a broad peak around Curie temperature, which is expeed in ferromagnetic materials. A slight change in slope is observed at low temperatures, which may be attributed to the antiferromagnetic transition observed at 16 K in the magnetization data. Additional low-temperature data would provide further clarity.

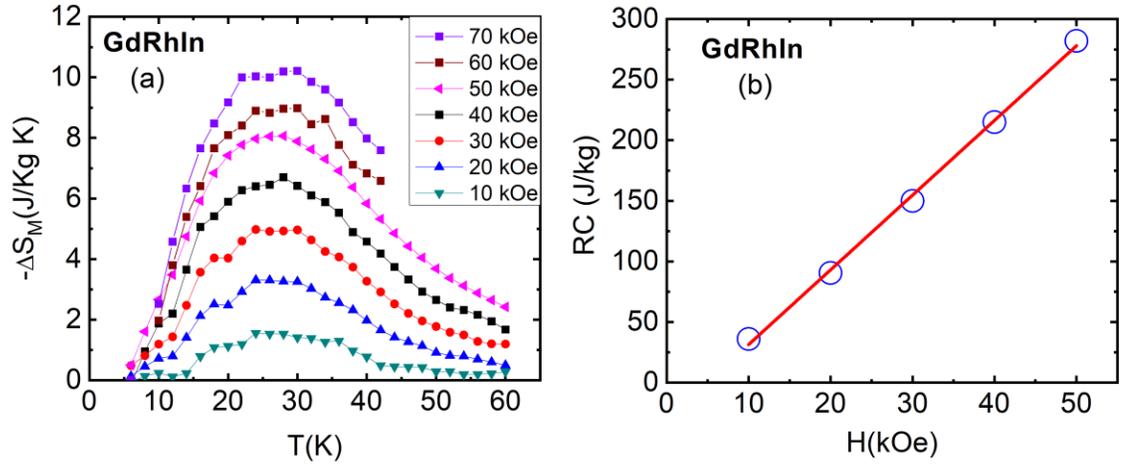

**Figure 4**. (a) The temperature dependence of isothermal magnetic entripy change, $\Delta S_M$ as a function of magnetic field for GdRhIn. (b) The field dependence of refrigerant capacity (RC) for GdRhIn compound.

In addition to $\Delta S_M$ and $\Delta T_{ad}$, another important parameter in respect to magnetic cooling efficiency of a materials is refrigerant capacity (RC) or relative cooling power (RCP), which can be calculaed as following [44]

$$RCP = \int_{T_1}^{T_2} \Delta S_M(T) dT \quad (5)$$

Where $T_1$ and $T_2$ represents the range of temperature at half maximum in $\Delta S_M$. In an ideal refrigeration cycle, RC is the measure of amount of the heat transferred from cold to hot reservoir. The RC values calculated from $\Delta S_M$ vs. $T$ plot and are shown in Fig. 4(b) for GdRhIn. It exhibits a linear dependence with magnetic field and is found to be 282 J/Kg for the field change of 50 kOe. RC could not be calculated for fields higher than 50 kOe due to a smaller number of data points. The value of RC is higher than or comparable to other Gd based rare earth intermetallics.

## 4. Conclusions

The GdRhIn was synthesized using arc melt technique. The crystal structure analysis shows hexagonal structure. GdRhIn shows two magnetic transitions at low temperatures. It shows antiferromagnetic to ferromagnetic transition at 16 K and ferromagnetic to paramagnetic transition at 34 K. Both the magnetic transitions are confirmed in the heat capacity data. The Debye temperature calculated from the heat capacity data are in range with other members of RTX family. The magnetocaloric effect calculated from the magnetization data was found to be larger than the other Gd based rare earth intermetallics, indicating GdRhIn to be a promising material for low temperature magnetic refrigeration.

**Acknowledgements**

The authors thank Prof. K. G. Suresh for the experimental support and fruitful discussion.